\documentclass{article}
\usepackage{spconf,amsmath,graphicx,amssymb,mathtools,commath}
\usepackage{tabularx}
\DeclareMathOperator*{\argmin}{arg\,min}
\usepackage{booktabs}

\title{Model-Based Deep Learning for Reconstruction of Joint k-q Under-sampled High Resolution Diffusion MRI}
\name{Merry P. Mani, Hemant K. Aggarwal, Sanjay Ghosh, and  Mathews Jacob\thanks{This  work  is  supported  by NIH 1R01EB019961-01A1.}}
\address{University of Iowa, Iowa, USA}
\begin{document}
%
\maketitle
\begin{abstract}
We propose a model-based deep learning architecture for the reconstruction of highly accelerated diffusion magnetic resonance imaging (MRI) that enables high resolution imaging. The proposed reconstruction jointly recovers all the diffusion weighted images in a single step from a joint k-q under-sampled acquisition in a parallel MRI setting. We propose the novel use of a pre-trained denoiser as a regularizer in a model-based reconstruction for the recovery of highly under-sampled data. Specifically, we designed the denoiser based on a general diffusion MRI tissue microstructure model for multi-compartmental modeling. By using a wide range of biologically plausible parameter values for the multi-compartmental microstructure model, we simulated diffusion signal that spans the entire microstructure parameter space. A neural network was trained in an unsupervised manner using an autoencoder to learn the diffusion MRI signal subspace. We employed the autoencoder in a model-based reconstruction and show that the autoencoder provides a strong denoising prior to recover the q-space signal. We show reconstruction results on a simulated brain dataset that shows high acceleration capabilities of the proposed method.\end{abstract}

\section{Introduction}
\label{sec:intro}

Diffusion weighted magnetic resonance imaging (DWI) is a widely used neuroimaging technique, which can provide rich information about a variety of tissue microstructural information including brain connectivity, and density of neurons \cite{Novikov2019}. The acquisition of diffusion MRI (dMRI) at high spatial resolution and on a large number of q-space points are needed to probe the tissue microstructural information and resolve the ambiguities in the parameters related to tissue microstructure\cite{Sotiropoulos2013,Jbabdi2015}. Conventional single-shot echo-planar (EPI) techniques have limited ability to improve the spatial resolution of dMRI. The long readouts required for higher resolution, often causes geometric distortions and blurring artifacts in the images. Several researchers have hence employed multi-shot EPI (msEPI) methods, where the k-space acquisition is segmented into multiple shots of shorter readout duration. However, a challenge with msEPI-based DWI acquisition scheme is the phase inconsistency between the shots. When the k-space data from the different shots are merged, these phase errors translate to ghosting artifacts in the images. Moreover, the multiple shots required to encode the images prolongs the acquisition time.

Several acceleration methods have been introduced in diffusion MRI to overcome the above challenges. These include (a) spatial (k-space) acceleration methods that rely on parallel MRI and compressed sensing \cite{Shi2015,Liao2017}, (b) q-space acceleration methods that acquire only a subset of the q-space data and rely on data priors to fill in the missing information\cite{Michailovich2011b,Welsh2013}, and (c) k-q acceleration methods that jointly under sample both k- and q spaces \cite{Mani2014,Schwab2018}. While the joint k-q under-sampling schemes can afford higher acceleration factors, the main challenges include (i) the high computational complexity of such scheme, resulting from the need to perform joint optimization, and (ii) the inability to account for complex diffusion models that do not conform with sparsity based models.

We propose a deep-learning based joint reconstruction algorithm for multi-shot diffusion MRI. The proposed scheme rely on a model based reconstruction that simultaneously performs phase correction and jointly recovers artifact-free DWIs from highly under-sampled acquisition. Specifically, a data fidelity term performs phase correction using the generalized SENSE reconstruction with known phase maps while a deep-learned prior exploits the redundancy in the q-space data. To achieve this, we trained a denoising auto-encoder (DAE) using training data generated by a generalized diffusion model. The non-linear network is shown to learn a projection to the data-manifold, thus denoising the images. We propose to use the residual error of the network, which can be used as a prior in a model-based reconstruction scheme. The reconstructed DWIs can then be used for further analysis to estimate the diffusion microstructure model parameters.

The proposed scheme has significant differences with deep-learning based q-space acceleration techniques \cite{Golkov2016}. This scheme rely on supervised learning to learn the mapping from the diffusion signal to the parameters of a specific model (e.g. NODDI) from fully sampled q-space images. By contrast, our focus is to recover the DWI data with high spatial and q-space resolution, which allows the fitting of any desired diffusion model. 
 \vspace{-1em}
\section{Methods}
\label{sec:format}

\subsection{Standard Multi-compartmental Diffusion Model}
The diffusion signal in the brain is often modeled by a multi-compartment model \cite{Novikov2019} that accounts for the intra-, and extra-neurite tissue compartments for each voxels, in addition to a isotropic compartment. The signal model is given by

\begin{equation}
 \rho(b, \mathbf g) = \rho_0\int_{\hat{ \bf{n}}}  {\cal{P}}(\hat{ \bf{n}}) \circledast K(b,\hat{\bf{g}} \cdot \mathbf n) ~d\hat {\bf{n}}
\label{model}
\end{equation}
where $\mathcal P$ is the fiber orientation distribution function (ODF) and $\circledast$ denotes a spherical convolution operation with a kernel $K $. The kernel is specified by 
 \begin{equation}
\hspace{0em} K(b,\zeta) = f_1e^{-bD_a\zeta^2} + f_2e^{-bD_e^{\perp} -b\left(D_e^{||} - D_e^{\perp}\right)\zeta^2}+f_{\rm iso} e^{-bD_{\rm iso}}, \notag
\end{equation}
where $f_{i}$'s are the volume fractions, $D$'s are the compartmental diffusivities, $b$ is the diffusion gradient strength, $\rho(b, \mathbf g) $ and  $\rho_0$ are the diffusion weighted and the reference non-diffusion weighted signal. The above diffusion signal model is very rich with several free model parameters. It has been noted to be useful for detailed microstructural analysis for the estimation of several tissue microstructure parameters when high quality diffusion data is available.

\subsection{Image Formation for msDWI }
Let $\mathbf \rho_{q}(x); q=1,..,Q$ represent the diffusion weighted image for the $q^{\rm th}$ location in q-space (the 3D space spanned by $b-\mathbf g$), where $\mathbf x$ represent the spatial co-ordinates. Then, the image acquisition model for an $S$-shot sampling in the presence of Gaussian noise $n$ can be represented as: 
\begin{equation}
  \label{eq:model1}
\mathbf {\hat y}_s = \mathcal{A}_s({\rho_{q}}) + \mathbf n, ~~ s=1:S
\end{equation}
where $\mathbf {\hat y}_s$ is the measured k-space data from shot $S$, and $\mathcal{A}=\mathcal{S}_s\circ \mathcal{F} \circ \mathcal {C} $. Here, $\mathcal{F}$, $\mathcal{S}_s$, and  $\mathcal{C}$ denotes Fourier transform, selection of the acquired k-space samples for a specific shot $s$, and weighting by coil sensitivities, respectively. For the phase compensated reconstruction for msDW data, we account the phase term into the coil sensitivity maps.
In a fully sampled scenario, the sampling patterns for the different shots are complementary; the combination of the data from the different shots will result in a fully sampled k-space. However, such fully sampled acquisitions result in long acquisition times. 

\begin{figure}
 \vspace{-0.5em}
~~~~\includegraphics[trim = 6mm 2mm 2mm 3mm, clip, width=0.2\textwidth]{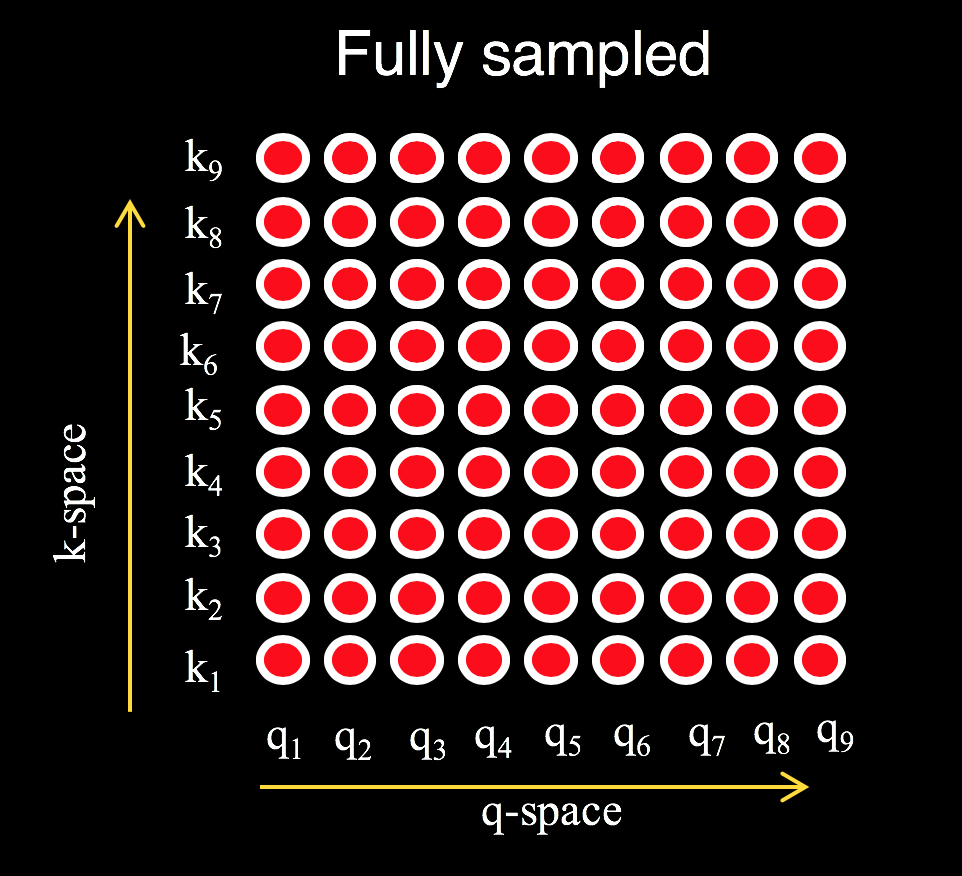}
\includegraphics[trim = 12mm 2mm 12mm 10mm, clip, width=0.2\textwidth]{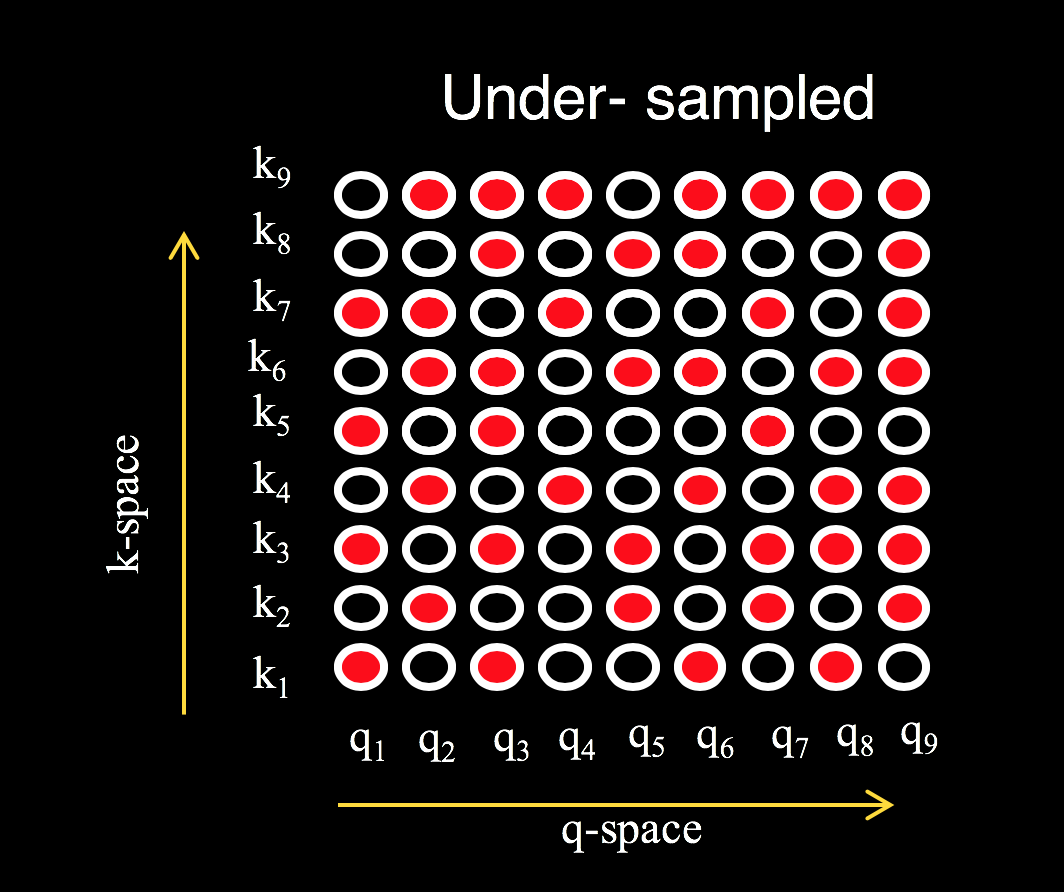}
\caption{Illustration of joint k-q undersampling. In the fully sampled k-q acquisition (left), all the k-space points corresponding to all the q-space points are sampled. In joint k-q under-sampling (right), the k-space samples for each of the q-space samples are randomly under-sampled.  }
\label{fig:fig1}
\vspace{-0.2em}
\end{figure}

 To simultaneously achieve high spatial and angular resolution using multi-shot sequences in a reasonable scan time, we propose to under-sample the joint k-q space of dMRI. Figure \ref{fig:fig1} represents the proposed joint k-q under-sampling that we pursue in the current work. This joint k-q acceleration scheme can be effectively achieved on the MRI scanners by randomly under-sampling the shots for each of the q-space sampling points.  
  We compactly denote the acquisition process as 
\begin{equation}
  \label{eq:compact}
 \widehat {\mathbf Y} = \mathcal{A}\left(\mathbf P\right) + \mathbf N,
\end{equation}
where $\mathbf Y$ is the Casoratti matrix (of dimension $ N_1 \times N_2 \times Q$), of the data corresponding to the different q-space points. 

\subsection{Model-based Joint Reconstruction Algorithm}

At high acceleration factors, the k-q under-sampled data needs to be jointly reconstructed. Denoting the k-space measurement matrix for the joint reconstruction as $\widehat{\mathbf Y}$, we propose to recover $\mathbf P$ by solving:
\begin{equation}
  \label{recon}
\mathbf P = \argmin_{\mathbf P} \norm{\mathcal {A}({\mathbf P})-\widehat{\mathbf Y }}_2^2  + \lambda ~~ \mathcal{R}({\mathbf P}).
\end{equation}
Here, the joint reconstruction enforces data consistency (DC) to the measured data using the generalized SENSE encoding operator $\mathcal{A}$ in the first term. The second term is an arbitrary regularization prior $\mathcal{R}$.
Priors including total variation spatial regularization and sparsity had been introduced by other researchers \cite{Michailovich2011b,Welsh2013,Mani2014,Schwab2018}. In our previous work \cite{Mani2014}, we employed sparsity priors, assuming a ball-and-stick diffusion dictionary model similar to MR fingerprinting. However, the extension of this idea  for the recovery of the parameters directly from the acquired data using fingerprinting-like recovery is complicated for diffusion models such as the model in Eq \eqref{model}. Evidently, the main challenge is the large size of the dictionary, resulting from the large number of free parameters, as well as the high coherence between the atoms that make $\ell_{1}$ minimization challenging. 

\vspace{-0.5em}
 \subsection{Denoising Autoencoder Prior}
 
We introduce a self-learning DMRI framework based on denoising autoencoders (DAE).  DAEs were introduced as unsupervised schemes to learn the data manifold. Theoretical results show that the DAE representation error is a measure of the derivative of the smoothed log density \cite{Vincent2008} of the data; the derivative is zero if the point is on the manifold, while it is high when the point moves away from the \emph{data-manifold}. 
Instead of using a dictionary based sparse prior, we propose to pre-learn a DAE from the dictionary $\mathbf Z$ such that:
\begin{equation}\label{daetraining}
\Theta^* = \arg \min_{{\Theta}} \mathbb E_{I} \left(\mathbb E_{\mathbf S \sim  \mathcal N(\mathbf 0,\sigma_i^2)}\|\mathcal D_{\Theta}\left(\mathbf Z+\mathbf S\right) - \mathbf Z\|_F^2\right)
\end{equation}
Here, $\mathbb E$ denotes the expectation operator and $\mathbf S$ is a noise realization with a zero mean complex Gaussian density with variance $\sigma_{I}^{2}$; the $\sigma_{i}$ are chosen from a set of variances, indexed within the set $I$. Once the parameters $\Theta$ are learned, we use the trained denoiser as a regularizer in plug-and-play framework \cite{zhang2017learning} in \eqref{recon} as:
\begin{equation}
\label{joint}
\mathbf P^{*} = \arg \min_{\mathbf P} \|\mathcal A (\mathbf P)-\widehat{\mathbf Y}\|^{2}_{2}+ \lambda~ \|\mathbf P - \mathcal D_{\Theta}(\mathbf P)\|^{2},
\end{equation}
where $\mathcal N_{\Theta}(\mathbf P) = \mathbf P-\mathcal D_{\Theta}(\mathbf P)$ is the DAE error. 
We solve the proposed joint recovery optimization using the alternating direction method of multipliers as follows:
\begin{eqnarray*}
\label{jointsteps}
	\mathbf P_{n+1} &=&  \arg \min_{\mathbf P} \|\mathcal A (\mathbf P)-\widehat{\mathbf Y}\|^{2}+ \lambda~ \|\mathbf P-\mathbf Q_n\|^2\\
	\mathbf Q_{n+1} &=& \mathcal D_{\Theta}(\mathbf P_{n+1}).
\end{eqnarray*}

  \subsection{Experimental Setup}
\subsubsection{Dictionary generation}

To generate the dictionary $\mathbf Z$, we employ the DWI signal model in Eq \eqref{model} and generate the diffusion signal $S(b, \mathbf g) $ for a range of model parameters. This model has 7 free parameters, all of which were varied within the physiological ranges to generate a dictionary that is a small subset of all possible diffusion signals. Specifically we used the ranges:  $f_{i}$'s $\in [0,1]$, and $D$'s $\in [0.1,3]$ \cite{Novikov2019}. The fiber direction $\hat{ \bf{n}}$ was varied for 30 different unit vectors in 3D space with crossing fibers simulated from the linear combination of these unit vectors. Since the reconstruction concerns the recovery of complex data, the generated signals were modulated with random phase terms, which counts as an additional parameter. 
 \vspace{-1em}
\subsubsection{DAE architecture and training}

 The generated diffusion signals were corrupted at various noise levels at $ $0$\%, $20$\%, $40$\%, \text{~and~} $60$\%$, and were used for training. The training data was fed to an autoencoder neural network. In this preliminary work, we employed an architecture with three fully connected layers, with ReLU activation function. The dimension of the input layer was the dimension of the q-space. The bottleneck layer was constrained to represent one fourth of the input dimension. 
 
\vspace{-1em}
\subsubsection{Testing data}
\begin{figure}
\vspace{-1em}
\centering
\includegraphics[trim = 12mm 12mm 12mm 6mm, clip, width=0.2\textwidth]{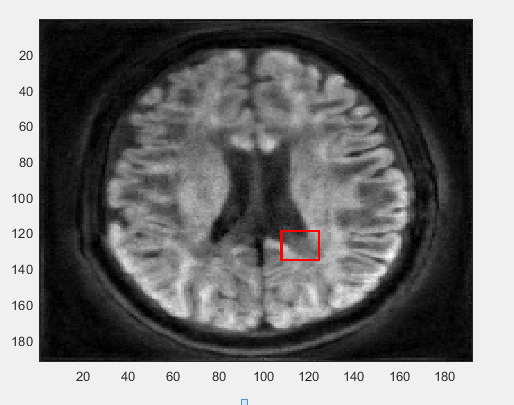}
~~~\includegraphics[trim = 1mm 1mm 1mm 1mm, clip, width=0.22\textwidth]{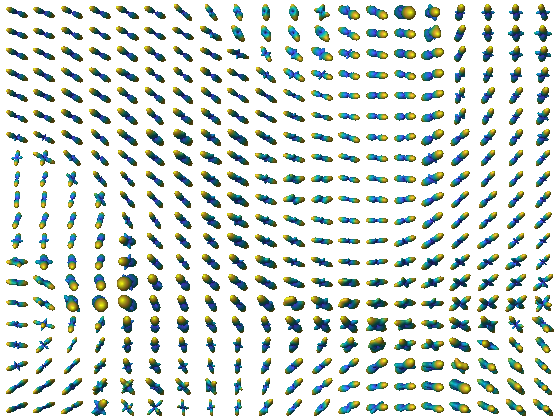}
\caption{The testing data is a brain data synthesized by estimating the ODFs from an in-vivo acquisition. The ODF (right panel) shown from the red boxed region show crossing fiber structures preserved.}
\label{fig:fsim}
\vspace{-1em}
\end{figure}

\noindent To test the joint reconstruction, we used a synthesized brain MRI data. This ground truth data was generated as follows: A high quality brain diffusion data was collected on a human volunteer using a variable density interleaved spiral acquisition with 22 spatial interleaves to achieve a high spatial resolution of 1.1mm in-plane. The data was collected on a 3T MRI with 8-channel head coil. 60 DWIs were acquired using the fully sampled spiral acquisition, which were independently reconstructed using CG-NUFFT SENSE reconstruction. The fiber orientation distribution functions in each pixel of this data was estimated and stored. These fiber orientation were used to generate the synthetic brain data. Figure \ref{fig:fsim} shows one DWI from this synthetic ground truth data, which displays crossing fibers in several voxels. 

The ground truth data was retrospectively under-sampled to generate the joint k-q under-sampled data for testing. Here, we assumed a Cartesian acquisition and the under-sampling was simulated using a multi-shot EPI scheme at different shot factors to study the various acceleration factors. Acceleration factors at R~=~4, 6, and 8 were considered. Random phase values were added to each of the shot images to simulate phase errors of the multi-shot imaging. 


\section{Results  }
\label{sec:pagestyle}

\begin{figure}[b]
\text{\rotatebox{90}{\fontsize{6}{2}\selectfont{{\hspace{2em} DAE outputs~~~~~~~~~~~ Input noisy DWIs  }}}}
\includegraphics[trim = 6mm 2mm 2mm 0mm, clip, width=0.45\textwidth]{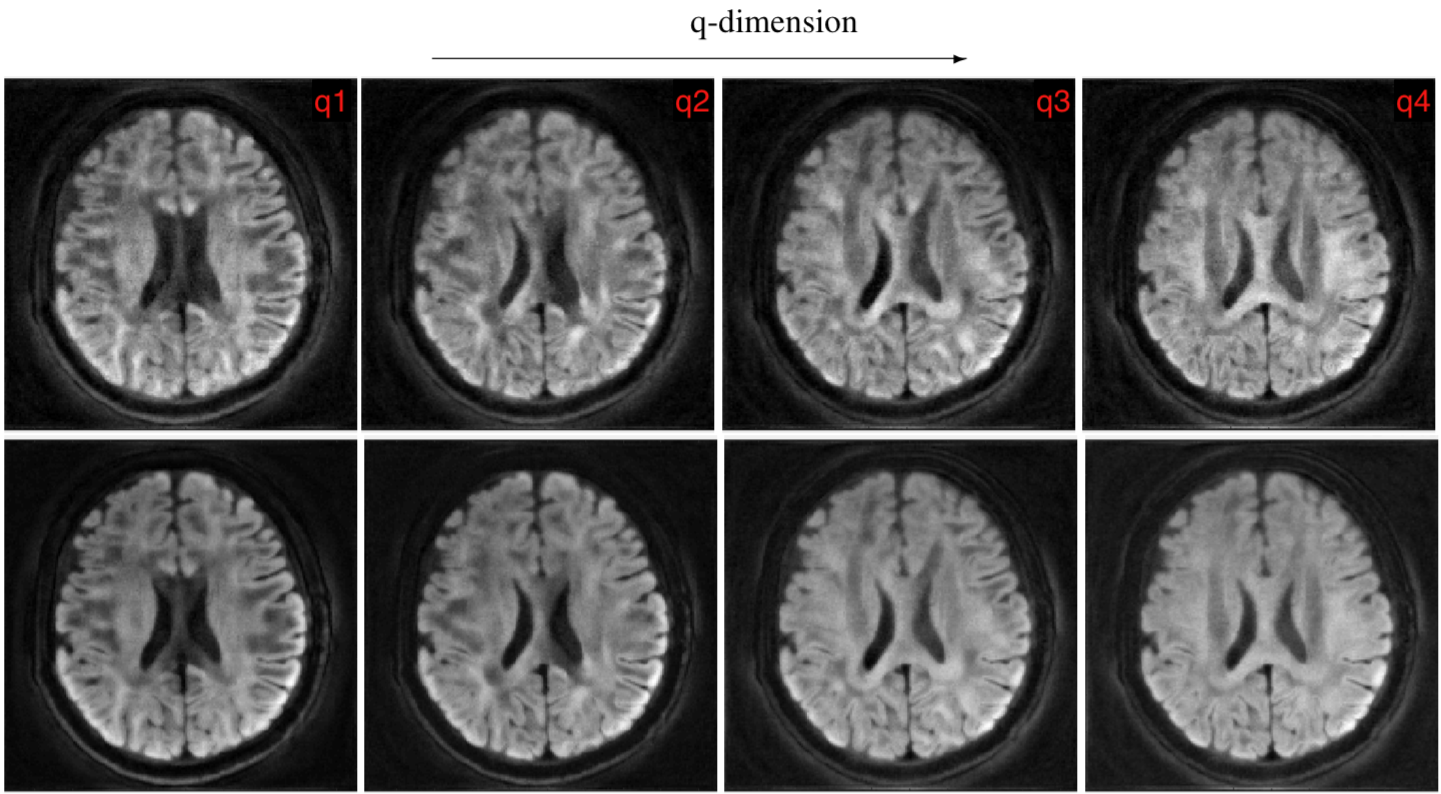}
\caption{The preservation of the diffusion contrasts along the q-dimension show that the manifold of q-space signals were learned by the DAE. }
\label{fig:denoise}
\end{figure}

The goal is to derive a regularization prior that can denoise the diffusion signal in the q-domain, which can be applied in a voxel-wise manner along the q-dimension during the joint reconstruction. Figure \ref{fig:denoise} shows the successful learning of the q-space signal manifold by the DAE. 
 The DAE was then used in the joint reconstruction in Eq \eqref{joint} to recover all 60 DWIs simultaneously for various undersampling factors following the alternating scheme discussed above.

\begin{figure}
 \vspace{-1em}
\text{\rotatebox{0}{\fontsize{6}{2}\selectfont{{\hspace{3em} $\mathcal{A}^H\widehat {\mathbf Y}   $ \hspace{4em}DC Iter 1 \hspace{5em}$D_{\Theta}$ Iter 1 \hspace{4em}$D_{\Theta}$ Iter 3 \hspace{4em}DC Iter 4 }}}}\\
\text{\rotatebox{90}{\fontsize{6}{2}\selectfont{{\hspace{3em} R=4 }}}}
\includegraphics[trim = 0mm 0mm 0mm 0mm, clip, width=0.49\textwidth]{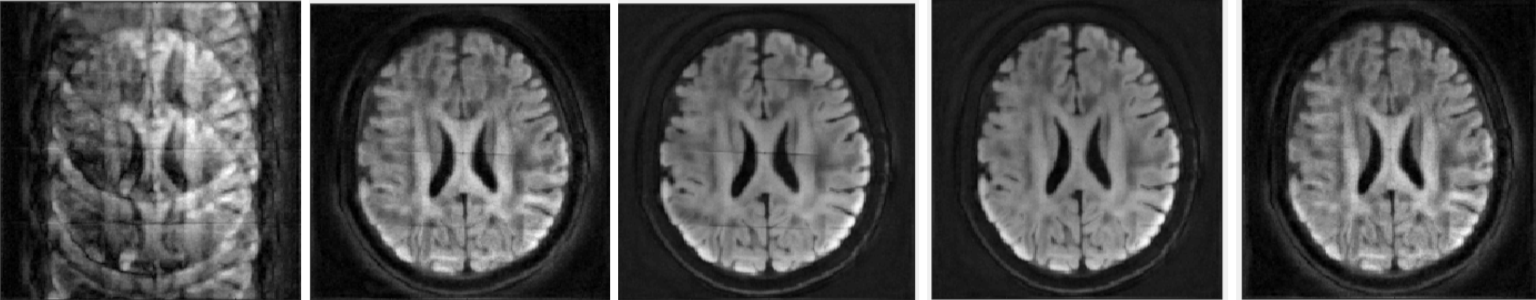}\\
\text{\rotatebox{90}{\fontsize{6}{2}\selectfont{{\hspace{3em} R=6 }}}}
\includegraphics[trim = 0mm 0mm 0mm 0mm, clip, width=0.49\textwidth]{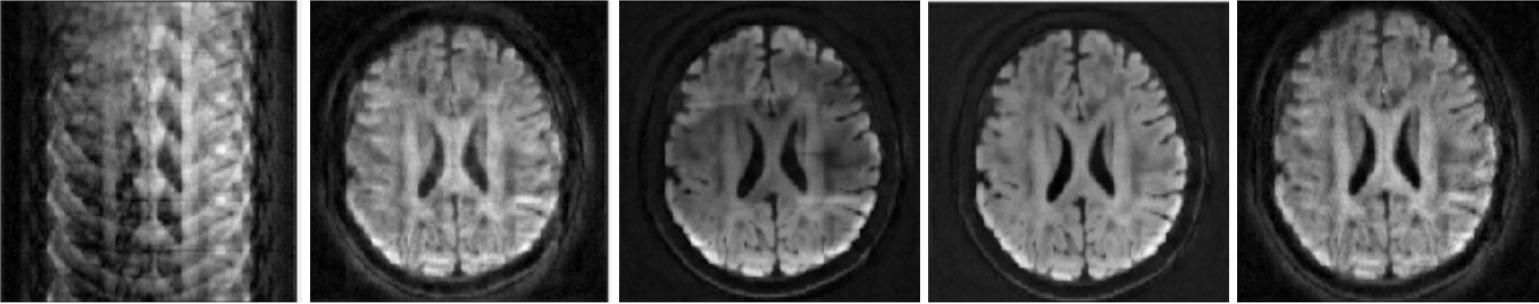}\\
\text{\rotatebox{90}{\fontsize{6}{2}\selectfont{{\hspace{3em} R=8  }}}}
\includegraphics[trim = 0mm 0mm 0mm 0mm, clip, width=0.49\textwidth]{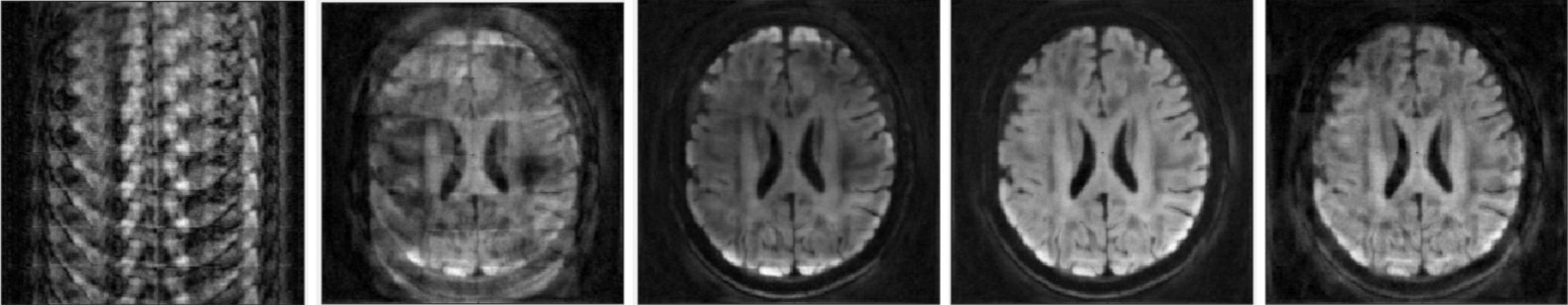}
 \vspace{-1em}
\caption{Joint reconstruction results shown for various under-sampling factors for a given diffusion direction. }
\label{fig:usfig}
 \vspace{-1em}
\end{figure}

 Figure \ref{fig:usfig} shows the results of the proposed reconstruction for various acceleration factors. Here, the first row shows the 4-shot case, where only one shot per DWI was sampled; the shot was chosen randomly for each DWI. Similarly, the second and third row show  6-shot and 8-shot cases. In all cases, only one random shot per q-space point was sampled. The performance of the denoiser at the first iteration as well as the DC updates at various stages of the reconstruction are shown in Figure 4. The root-mean-square error (RMSE) and peak signal-to-noise ratio (PSNR) for various acceleration factors are reported in Table 1. It is clear from Figure \ref{fig:usfig} and Table 1 that the proposed DAE regularizer is an efficient recovery prior for the reconstruction of highly under-sampled data.

\begin{table}
	\centering
	\caption{Reconstruction error of the proposed scheme for various undersampling factors.} ~\\
	\label{tab:time}
	\begin{tabular}{| c | c | c|} \hline
		Acceleration     & RMSE & PSNR    \\ \hline
		$R = 4$    & $0.0176$  & $35.04$        \\ \hline
		$R = 6$    & $0.0548$  & $25.19$        \\ \hline
		$R = 8$    & $0.079$  & $22.01$        \\ \hline
	\end{tabular}
\end{table}

\section{Discussion \& Conclusion}

We introduced a model based deep learning framework for the
joint recovery of DWIs from joint k-q under-sampled data. In this preliminary work, we show the feasibility of employing a DAE to prelearn the projection to q-space signal manifold. The prelearning was performed using simulated diffusion data using a general diffusion model with several degrees of freedom. We note that the accuracy of the DAE is determined by the training data; specifically, the more range of parameters used to simulate the data will result in improved denoising. The need to account for multiple fiber orientations per voxel significantly inflate the parameter space. In the current study, we only considered 30 unique fiber directions, which may have contributed to reduced accuracy. In future work, we would explore the scenario with larger dictionary with more fiber directions. We also plan to extend this work for the recovery of multi-shell dMRI data in the future.\\

\vspace{-2em}
\bibliographystyle{IEEEbib}
\small
\bibliography{Dmri_HD.bib}

%
%
%
%
%
%

\end{document}